\documentclass{emulateapj}

\usepackage{epsfig}
\usepackage{apjfonts}
\usepackage{epsfig}
\usepackage{rotating}

\newcommand{\chan}{{\sl Chandra}}

\def\aciss3{{ACIS-S3}}

\begin{document}

\submitted{Submitted to ApJ 2002 November 27; accepted 2003 March
26; to be published 2003 July 10}

\title{
The Variable Jet of the Vela Pulsar.}

\author{
G.\ G.\ Pavlov, M.\ A.\ Teter, O.\ Kargaltsev, and D.\ Sanwal}
\affil{The Pennsylvania State University, 525 Davey Lab,
University Park, PA 16802, USA} \email{pavlov@astro.psu.edu}

\begin{abstract}

Observations of the Vela pulsar-wind nebula (PWN) with the
{\sl Chandra X-ray Observatory} have revealed a complex,
variable PWN structure,
including inner and outer arcs, a jet in the direction of the pulsar's
proper motion, and a counter-jet in the opposite direction, embedded
in diffuse nebular emission. The jet consists of a bright, $8''$-long
inner jet, between the pulsar and the outer arc,
and a dim, curved outer jet that
extends up to $\sim 100''$ in approximately the same direction.
 From the analysis of thirteen {\sl Chandra} observations
spread over $\approx 2.5$ years
we found that this outer jet shows particularly strong variability,
changing its shape and brightness. We observed bright blobs in
the outer jet moving away from the pulsar with apparent speeds
(0.3--$0.6)\, c$ and fading on time-scales of days to weeks.
If the blobs are carried away by a flow along the jet,
the observed variations suggest mildly relativistic
flow velocities,
about (0.3--$0.7)\, c$.
The spectrum of the outer jet
fits a power-law model with a photon index $\Gamma=1.3\pm0.1$.
For a distance of 300 pc, the apparent
average luminosity of the outer jet in the 1--8 keV band
is about $3\times 10^{30}$ erg s$^{-1}$,
compared to  $6\times 10^{32}$ from the whole PWN within $42''$ from the pulsar.
The X-ray emission of the outer jet
can be interpreted as synchrotron radiation of
ultrarelativistic electrons/positrons.
This interpretation allows one to estimate the magnetic field,
$\sim 100$ $\mu$G, maximum energy of X-ray emitting electrons,
$\sim 2\times 10^{14}$ eV, and
energy injection rate, $\sim 8\times 10^{33}\, {\rm erg\, s}^{-1}$,
for the outer jet.
In the summed PWN image, we see a faint, strongly bent, extension of the outer jet.
This bending could be caused by combined action of a wind within the supernova
remnant, with a velocity of a few $\times 10$ km s$^{-1}$,
along with the ram pressure
due to the pulsar's proper motion. The more extreme bends closer to the pulsar,
as well as the apparent
side motions of the outer jet, can be associated with kink instabilities
of a magnetically confined, pinched jet flow.
Another feature found in the summed image is a dim, $\sim2'$-long
outer counter-jet, which also shows a power-law spectrum with
$\Gamma \approx 1.2$--1.5.
Southwest of the jet/counter-jet (i.e., approximately perpendicular
to the direction of pulsar's proper motion), an extended region of
diffuse emission is seen.
Relativistic particles responsible for this radiation are apparently
supplied by the outer jet.

\end{abstract}


\section{Introduction}
Thanks to the outstanding angular resolution of the X-ray telescopes aboard the
{\sl Chandra X-ray Observatory}, we are able to study
the structure of pulsar-wind nebulae (PWNe), formed by
relativistic outflows from pulsar magnetospheres.
Particularly detailed {\sl Chandra} images have been obtained
for PWNe around the Crab pulsar (Mori et al.\ 2002; Hester et al.\ 2002), Vela
pulsar
(Helfand, Gotthelf, \& Halpern  2001;
Pavlov et al.\ 2001a), and PSR B1509--58 (Gaensler et al.\ 2002).
Each of these images shows an approximately
axially-symmetric PWN morphology,
with  an extended structure stretched along the symmetry axis.
This suggests that such structures, usually called {\em jets},
are common to at least young pulsars.
Most likely, these jets are associated with collimated outflows
of relativistic particles along the pulsar's rotation axes.
Investigations of pulsar jets allow one to understand
the origin and properties of pulsar winds and their interaction
with the ambient medium. Moreover, since jets have been observed in
many astrophysical objects (e.g., AGNs, Galactic microquasars),
studying pulsar jets may shed light on the mechanism of jet formation
in these objects as well.

Due to its proximity ($d\simeq 300$ pc, from the annual parallax
of the Vela pulsar --- Caraveo et al.\ 2001; Dodson et al.\ 2003a),
the Vela PWN is particularly well suited
for studying the pulsar outflows.
This PWN was discovered with {\sl Einstein} (Harnden et al.\ 1985)
and studied extensively with {\sl ROSAT}
and {\sl ASCA} (Markwardt \& \"Ogelman 1998,
and references therein).
The {\sl Einstein} and {\sl ROSAT}
observations with the High Resolution Imagers (HRIs) have established
the overall kidney-bean shape of the Vela PWN, with a symmetry axis
approximately co-aligned with the pulsar's proper motion.
The first observations of the Vela PWN with {\sl Chandra}
(Pavlov et al.\ 2000; Helfand et al.\ 2001) have shown
a detailed structure of the nebula, including
inner and outer arcs, northwest and southeast jets, and bright knots
in the southeast part of the PWN.
Furthermore,
Pavlov et al.\ (2001a) found that the nebula elements were variable
in position, brightness, and, perhaps, spectrum.
Based on two observations with the {\sl Chandra}'s
Advanced CCD Imaging Spectrometer (ACIS), Pavlov et al.\ (2001a)
 noticed
a long, curved extension of the northwest jet
in approximately the same direction, well beyond the apparent
termination point of the bright ``inner jet'' at its intersection
with the outer arc. This dim ``outer jet'' (dubbed ``filament'' in that
paper) had a brighter ``blob'' close to the jet's end, which showed
particularly strong displacement between the two observations
taken 7 months apart.

To understand the nature of the variability of the Vela PWN,
we carried out a series of eight monitoring observations with the
{\sl Chandra} ACIS. These observations
 have confirmed the dynamical
structure of the PWN, with most dramatic changes occurring in the
outer jet. Moreover, we were able to detect an ``outer counter-jet'',
a much dimmer extension
of the southeast (counter-)jet, which is clearly seen only in the summed
image, together with highly asymmetric diffuse emission.
In the present paper, we concentrate
mainly on the highly variable outer jet, which has been detected
in ten ACIS observations and three observations with the
High Resolution Camera (HRC),
deferring a detailed analysis of the other PWN elements to a future
work.

We describe the observations, their analyses, and observational results in \S2,
and discuss possible interpretations in \S3. The summary of our results and
concluding remarks are presented in \S4.
\section{Observations and Data Analysis}
\subsection{ACIS observations}
Series of ten ACIS observations of the Vela pulsar and its PWN were carried
out from 2000 April 30 through 2002 August 6.
The dates of the observations
and the exposure times are given in Table \ref{obs}. In  all
observations the target was imaged on the back-illuminated chip
S3,  less affected by the radiation-induced changes in CCD charge transfer
efficiency than the front-illuminated chips.
To image the whole
PWN on one chip, the pulsar was offset from the ACIS-S aim-point by
$-1\farcm 5$ along the chip row in each observation.
In the first two observations only half of the chip was read out
(1/2 Subarray mode), with a frame time of 1.5 s, to reduce the pile-up
from the bright pulsar.
The rest of the observations  were done in the Full Array  mode, with
a frame time of 3.24 s.
  The  telescope focal
plane temperature for all ACIS observations was
$-120$~C. We used the CIAO
software\footnote{Chandra Interactive Analysis of Observations --- see
{\tt http://cxc.harvard.edu/ciao/}.}, v.2.2.1 (CALDB 2.10) for the data analysis,
proceeding from Level 1 event files.
The energy and grade of each event were corrected for
charge transfer inefficiency (CTI) using the tool developed by Townsley et al.\ (2000).

To examine the changes in the PWN surface brightness,
we have scaled the number of counts
per pixel
 for each of the ten ACIS observations
as $N_i^\prime = N_i (t_5/t_i)$, where
$t_i$ is the
effective exposure time of $i$-th observation,
and $t_5=17933$ s
is the exposure time of the reference observation (ObsID 2813).
 In addition, we restricted the energy range to 1.0--8.0 keV
to reduce the background contribution at higher energies
and get rid of the trailed images (readout streaks)
produced by
the bright pulsar, whose spectrum is softer than that of the PWN.
To produce adaptively smoothed images, we use the CIAO tool {\tt csmooth},
with the smoothing scale such that the signal-to-noise ratio is between 4 and 5.

Applying the CIAO
tool {\tt wavdetect} to the individual images, we found 15 point-like
sources on the S3 chip; 9 of them,
including the Vela pulsar,
we seen in at least
4 images.
The rms deviation
of the {\sl Chandra} pipe-line coordinates of 8
of the 9
sources is $\le 0\farcs7$,
while for one off-axis source it is
about $1\farcs4$.
This source is likely extended because {\tt wavdetect}
finds two sources in some of the images.
Therefore, we exclude this
source and conclude that the astrometry is better than
$0\farcs7$.
We will use this value
as the uncertainity
of the coordinate measurements.

\subsection{HRC observations}
To increase the total time span and the sampling, we also included
three archival {\sl Chandra}\ HRC observations in our analysis
(see Table 1 for the ObsIDs and exposure times).
In all the three observations the Vela
PWN was imaged on the HRC-I (imaging) plate.
Pipe-line processed
Level 2 event files from the archive were used for the analysis,
without further reduction.
We scaled the binned images
to the 2000 January 20 exposure time and adjusted the color
scale
to reveal the similar amount of detail that was seen in
the ACIS images.
To produce the smoothed images, each image
was binned at a 2 pixel by 2 pixel scale and then smoothed similar to
the ACIS images.

\begin{table}[h!tb]
\label{powertable} \caption[]{
        Summary of the observations. \label{obs}
        }
\begin{center}
\begin{tabular}{llccc} \tableline\tableline
 Panel & Obs ID & Instrument &  Date of observation & Exposure time (s)   \\  \tableline
1&   1518   & HRC-I &   2000-01-20 &  49,765 \\
2&    364   & HRC-I &   2000-02-21 &  48,050 \\
3&        128   & ACIS-S &  2000-04-30 &  10,577  \\
4&        1987  & ACIS-S &  2000-11-30 &  18,851  \\
5&        2813  & ACIS-S &  2001-11-25 &  17,933    \\
6&        2814  & ACIS-S &  2001-11-27 &  19,870   \\
7&        2815  & ACIS-S &      2001-12-04 &  26,960  \\
8&        2816  & ACIS-S &  2001-12-11 &  18,995   \\
9&        2817  & ACIS-S &      2001-12-29 &  18,938    \\
10&   1966  & HRC-I  &  2002-01-13 &  49,464 \\
11&       2818  & ACIS-S &      2002-01-28 &  18,663   \\
12&       2819  & ACIS-S &      2002-04-03 &  19,920    \\
13&   2820  & ACIS-S &  2002-08-06 &  19,503 \\ \tableline
        \end{tabular}
   \\ \rule{0mm}{5mm}

\end{center}
\end{table}
\subsection{Spatial structure}
The large-scale X-ray structure of the Vela PWN is shown in the summed
ACIS image of observations 2813--2820 in Figure 1. The brightest,
kidney-shaped part of
the PWN (white and yellow in Fig.\ 1) is approximately symmetric with respect
to the direction of the pulsar's proper motion (PA $\simeq 307^\circ$,
in the Local Standard of Rest  ---
see Caraveo et al. 2001, and references therein). This deep image
clearly reveals a long ($\approx 1\farcm 7$, as measured from the
pulsar) outer jet, approximately in the direction
of the proper motion, which is somewhat curved and brightened at its end.
A much fainter outer counter-jet is seen, for the first time,
in the opposite direction. Finally, we see highly asymmetric
diffuse emission feature southwest of the jet/counter-jet line, which is
obviously connected to the ``main body'' of the PWN.

\begin{figure}
\begin{center}
\includegraphics[width=4.3in]{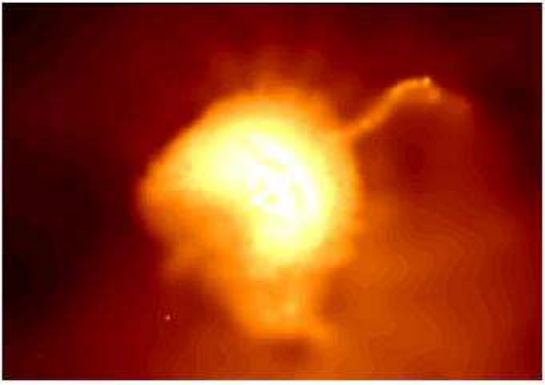}
\vspace{-3.0cm} \figcaption{Merged smoothed image, $4\farcm6
\times 3\farcm6$, of the Vela PWN scaled to show the outer jet and
outer counter-jet (northwest and southeast of the bright PWN core,
respectively), and the faint diffuse emission southwest of the
core. In all images in this paper North is up and East to the
left. \label{zeroth}}
\end{center}
\end{figure}

Choosing different brightness scale and color scheme,
one can resolve a fine structure within the bright core of the PWN.
It is depicted in Figure 2 which shows a zoomed individual
image of 2002 January 28.
The most apparent elements of this structure,
with the pulsar [1] at the center, are the inner arc
[2],
the outer arc [3], the inner jet [4], and the inner counter-jet [5].
The inner jet is directed northwest from the pulsar (PA $\simeq 307^\circ$)
in the direction of the pulsar's proper motion,
and the
counter-jet is directed toward the southeast (PA $\simeq 127^\circ$
--- Pavlov et al.\ 2000).
The bright core (red and yellow in Fig.\ 2) is surrounded by
a ``shell'' [6] of diffuse emission (green in Fig.\ 2).
The outer jet [7] looks like an extension of the much brighter inner jet,
well beyond the apparent termination point of the inner jet at its
intersection with the outer arc.
It is clearly seen
on this image, contrary to the much dimmer outer counter-jet.
Comparison of the individual images shows that all the PWN elements
are variable, especially the outer jet.
Throughout this paper, we will concentrate our analysis upon the
dynamic outer jet,
though the inner jet and counter-jet will be considered for comparison.

\begin{figure}
\begin{center}
\includegraphics[width=4.3in]{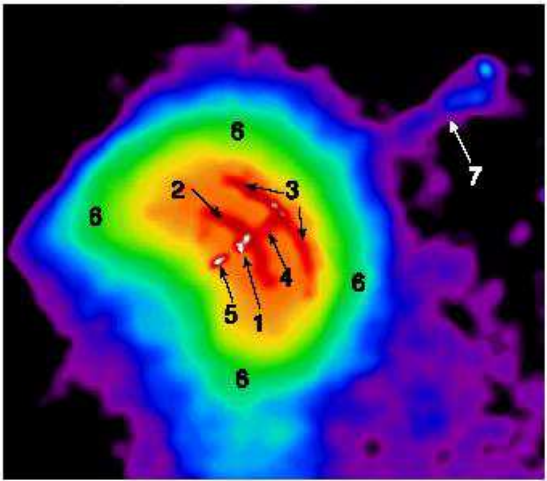}
\vspace{-1.5cm}
 \figcaption{\chan\ ACIS-S3 image, $2\farcm1 \times 2\farcm7$, of
the Vela PWN sh owing the structural elements: (1) the Vela
pulsar,  (2) the inner arc,  (3) the outer arc  (4) the inner jet,
(5) the counter jet,  (6) the shell, (7) the outer jet.
\label{first}}
\end{center}
\end{figure}

The outer jet
extends away
$\approx 0.14\, d_{300}$ pc ($4.3\times
10^{17}d_{300}\,\, {\rm cm} = 0.45\,d_{300}$ lt-yr) from the pulsar,
where $d_{300}$ is
the distance to the pulsar in units of 300 pc.
The characteristic width (diameter) of the outer jet is about
$3 \times 10^{16}d_{300}$ cm.
Because the outer jet is generally co-aligned with the
inner jet, it is natural to assume that the inner and outer jets
are connected at the intersection with the outer arc,
although the dim outer jet can hardly be seen within the shell.
Figure 2
shows a typical view of the outer jet structure.  It
is not as straight as the inner jet; in this image it slightly bends
southward from the inner jet/counter-jet direction,
and back to the north near the end.
Other individual images
(Figs.\ 3 and 4)
show  different, sometimes more extreme,
bending
(e.g., panel 2 in Fig.\ 3 and panel 13 in Fig.\ 4).

The average (background-subtracted) surface brightness
of the outer jet
in the ten ACIS images is about 0.05 counts arcsec$^{-2}$ ks$^{-1}$, in the 1--8 keV band,
with extremes of variablilty from $0.032\pm0.004$ to $0.060\pm0.003$, in the same units
(errors here and below are 1 $\sigma$ errors).
It is seen from Figures 1 and 2, that the outer jet brightness varies
along its length. Choosing yet another brightness scale
(Figs.\ 3 and 4), we resolve
compact regions of enhanced emission within the outer jet (blobs, hereafter),
with brightness
up to $0.16\pm0.02$ counts arcsec$^{-2}$ ks$^{-1}$,
typically 2--3 times brighter than
 the remainder of the outer jet.
In turn, the
outer jet is approximately
3 times brighter than the background on the northeast side of the jet,
with brightness $0.021\pm0.002$ counts arcsec$^{-2}$ ks$^{-1}$.
The background on the southwest side of the outer jet is about 50\%
brighter than that on the northeast side.
This difference is quite clearly seen from Figure 1, which also shows
that the increased background is
connected to a region of enhanced
diffuse emission on
the southwest side of the bright PWN.
The average surface brightnesses of the inner jet and counter-jet
are 1.5 and 1.8
counts arcsec$^{-2}$ ks$^{-1}$,
respectively, ranging from 1.1 to 1.9 for the inner jet and
from 1.3 to 2.8 for the counter-jet, with typical errors of
0.2 counts arcsec$^{-2}$ ks$^{-1}$.
The background-subtracted surface brightness of the outer counter-jet
is $\approx 0.007$
counts arcsec$^{-2}$ ks$^{-1}$,
as measured from the summed image in Figure 1.

\begin{figure*}
\begin{center}
\includegraphics[width=4in]{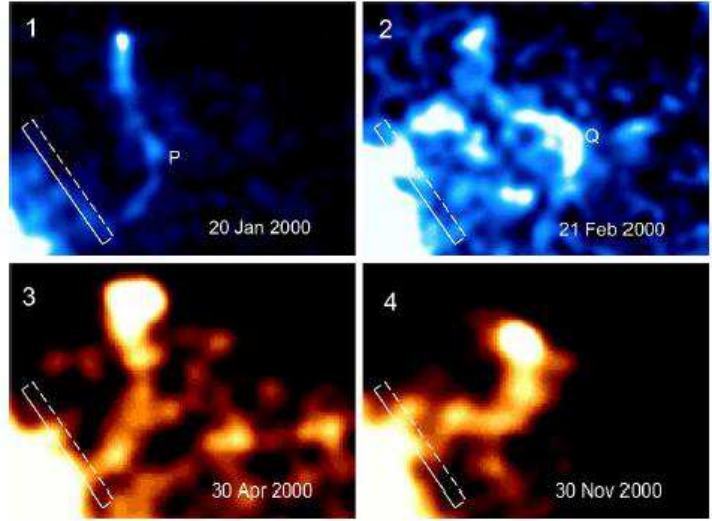}
\figcaption{Images of the outer jet from the observations of year
2000. The HRC-I and ACIS-S3 images are in blue and orange,
respectively. The panels are numbered in accordance with Table 1.
The size of each panels is $73''\times 53''$.
  The boxes, $28''\times2\farcs6$, at same sky position
in all the panels, are overplotted to guide the eye.
 The points P and Q mark westward extremes of the bent jet
in the two HRC observations; they were used to estimate the jet's
motion between the two observations (see text for details).
\label{2000}}
\end{center}
\end{figure*}

\begin{figure*}
\begin{center}
\includegraphics[width=6in]{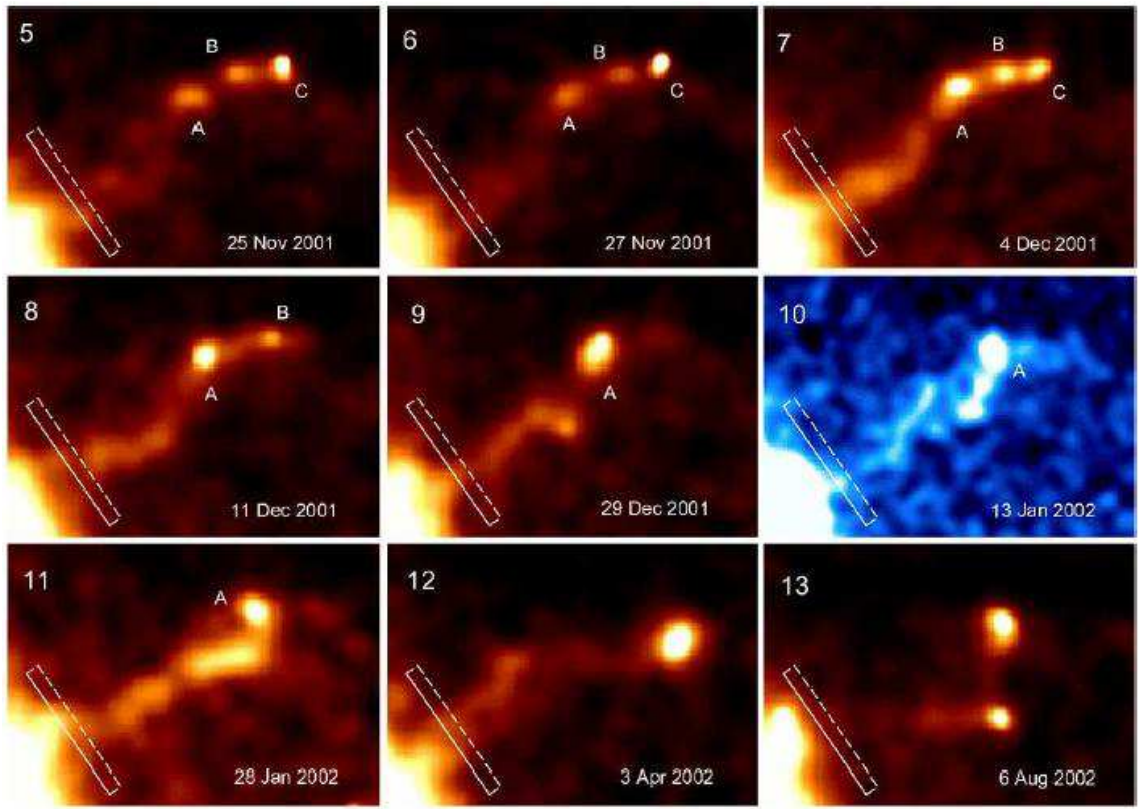}
\figcaption{Sequence of \chan\ images of the outer jet observed in
2001--2002.  The
 $73''$ by $53''$ panels are numbered according to Table 1.
The letters A, B, and C label the moving bright blobs identified
in observations 5 through 11.
  The boxes, $28''\times2\farcs6$,  at same sky position in all the panels,
were used to demonstrate the jet's sideways motions (see Fig.\
\ref{projections}). \label{details}}
\end{center}
\end{figure*}
%
\subsection{Spatial variation}
Figures 3 and 4
give the time sequence of the \aciss3 and HRC-I observations for
the outer jet.  Each of the panels corresponds to the appropriate data set
from Table 1. The variability of the outer jet is clearly seen,
in both the HRC and ACIS images.
Over the thirteen observations, with different
periods of time between each of them, we distinguish three different
types of variability in the outer jet.
First, the outer jet shifts from side to side, bending and
apparently twisting.  Second, the blobs move outward along the
outer jet.  Finally, the blobs change in brightness and eventually disappear.

A very vivid example of
the structural change
in just one month is demonstrated by the first two {\sl Chandra}
HRC observations
(panels 1 and 2 of Fig.\ 3).
In panel 1, the outer jet looks like a ``hook''
attached to the bright spot on the outer arc, with
the outer end of the hook pointing due north and terminating
at an elongated blob. In panel 2, the hook evolved into a
much more bent loop-like structure, which terminates with
apparently the same (albeit fainter) blob at approximately
the same location.
The structure has moved between two observations, indicating that
the sideways shift (e.g., from point P in panel 1 to point Q of panel 2) is
approximately $12''$ in the 32 days separating the observations.  This
translates to a
sideways speed of 0.6 c
(all the speeds in this section are given for $d=300$ pc).
Another example is presented by the first two ACIS images
(panels 3 and 4)
separated by 7 months.
The shape of the outer jet has changed from
  pointing due north and
terminating with a blob
to an S-shaped
structure.
The displacement of the blob (which easily could be another blob given that the
time between the observations is so long)
suggests a speed $\ga 0.1 c$.
In observations 5--11 (Fig.\ 4)
the jet appears almost straight, with
a slight southward curve.  The short time
(16 days) between the fifth and eighth
observations suggests that the sideways shifts of the jet
occur on
 a time
scale of order of weeks.
A dramatic change of the jet shape is again seen
in the last observation 13, separated by 4 months
from the previous observation. In the last image,
the jet points due west beyond the PWN shell and
then abruptly turns to the north, with a bright blob
at the turning point.
Overall, the
  small-scale structural changes are seen in all of the
observations.  For instance, the ``base'' of the outer jet
(where it leaves the shell
--- see the white boxes in Figs.\ 3 and 4)
shifts from one observation to the next.
Figure 5 shows
smoothed one-dimensional distributions of counts across the outer jet
(i.e., along the length of the narrow boxes in Fig.\ 4) in eight consecutive
ACIS observations. Typical apparent speeds of these shifts are of order
a few tenths of speed of light.

 The apparent speeds of the blobs along the jet are of
similar magnitude. The proper motions of the blobs A and B seen in
several panels of Figure 4 are shown in Figures 6 and 7. The $1''$
uncertainty of the position of the blobs is obtained by adding in
quadrature the 0\farcs7 uncertainty from the astrometry and an
additional uncertainty, of approximately the same value, from
measuring the centroid of the blob from the smoothed images. Blob
B vanishes in panel 9, so its apparent speed is calculated from
four observations. The apparent speeds of blobs A and B are
$(0.35\pm 0.06)\, c$ and $(0.51\pm 0.16)\, c$, respectively. Blob
C vanished  quickly, having an (unconstrained) apparent speed of
$(0.6\pm 0.7)\,c$. Assuming that the measurements of the apparent
velocities of the blobs inside the outer jet give the bulk flow
speed, $v_{\rm flow} \sim (0.3$--$0.7)\, c$, and the jet does not
deviate strongly from the sky plane, we obtain $t_{\rm
flow}=l_{\rm jet}/v_{\rm flow} \sim (0.6$--$1.5)$ yr for the
travel time along the jet. The actual travel time may be
significantly different if the jet direction strongly deviates
from the sky plane and relativistic effects are important (see
\S3.1).

\begin{figure}
\begin{center}
\includegraphics[height=6in]{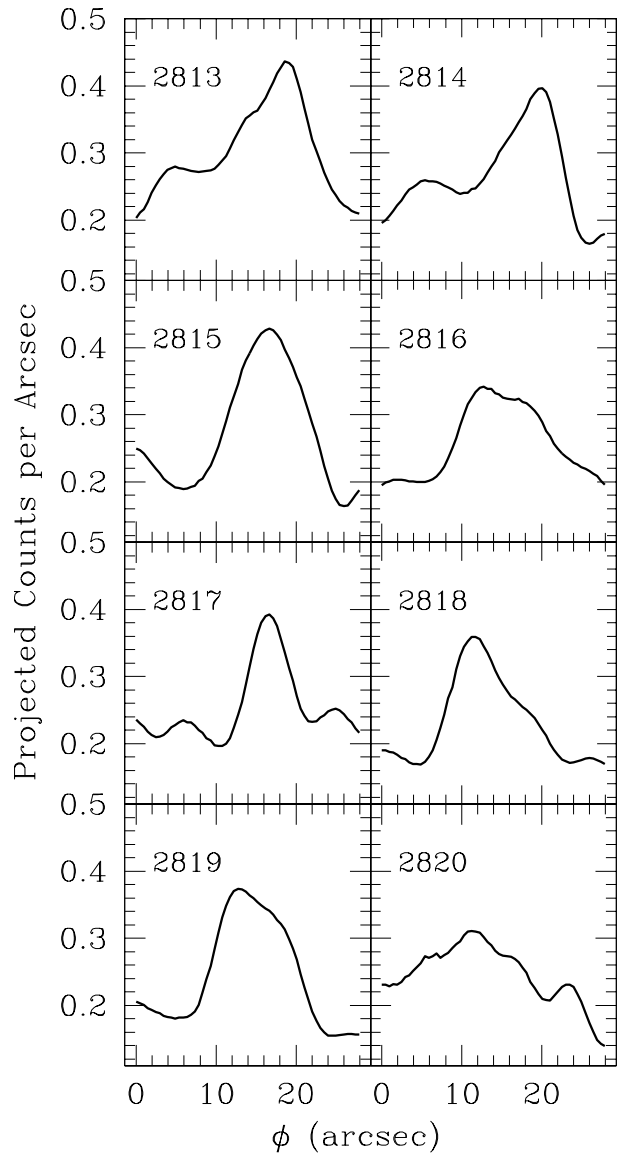}
\figcaption{Smoothed distributions of counts across the outer jet
at a distance of $50''$ from the pulsar in eight ACIS observations
(ObsID 2813--2820). The number of counts (per arcsecond)
integrated along the short dimension of the
 boxes
in Figure \ref{details} is plotted as a function of coordinate
$\phi$ along the box length. \label{projections}}
\end{center}
\end{figure}

\begin{figure}
\begin{center}
\includegraphics[width=3.4in]{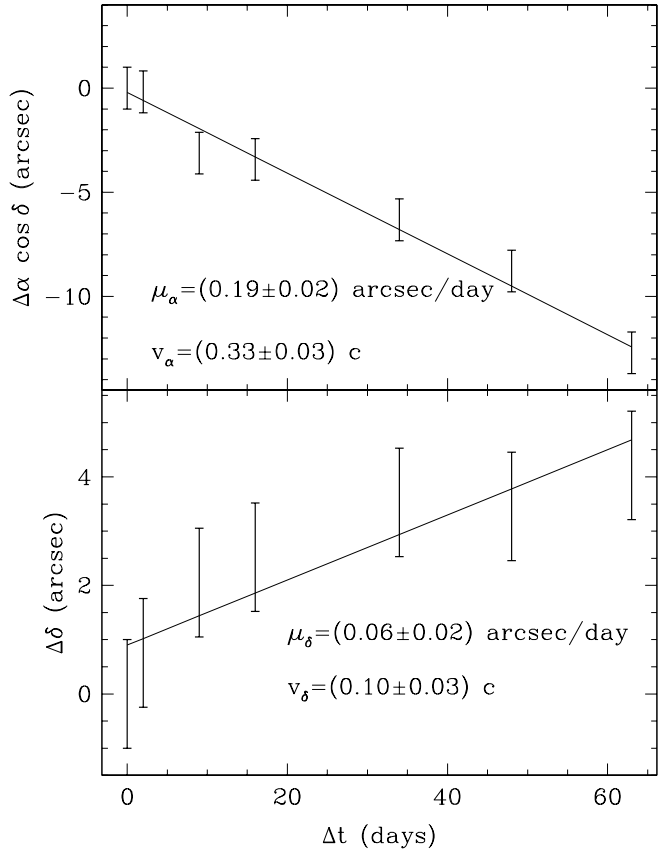}
\figcaption{Proper motion of blob A in panels 5--11 of Figure
\ref{details}. The upper and lower panels show the motions along
the right ascension and declination, respectively. The straight
lines are the least-square fits assuming constant speeds. Under
this assumption, the apparent speed of blob A is ($0.35\pm0.06)\,
d_{300}\, c$. \label{knotA}}
\end{center}
\end{figure}

The blobs change brightness as they move and
disappear.  For
instance, blob C  slightly brightens
in two days between observations
5 and 6, then fades somewhat between observations 6 and 7,
and disappears completely
after another week
(by observation 8).
Similarly, the brightness of blob B changes
noticeably through observations 5--8, while this blob vanishes in
observation 9.
No such blobs
are found in either the inner jet or the counter-jet.
\begin{figure}
\begin{center}
\includegraphics[width=3.4in]{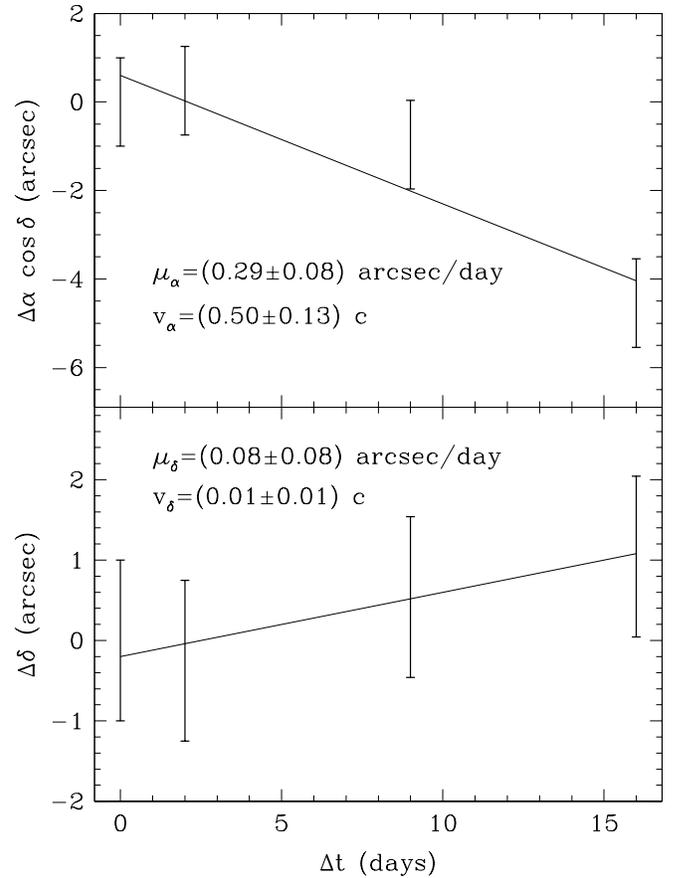}
\figcaption{Proper motion of blob B in panels 5--8 of Figure
\ref{details}.  The apparent speed of blob B is ($0.51\pm0.16)\,
d_{300}\, c$. \label{knotB}}
\end{center}
\end{figure}

\subsection{Spectrum and luminosity}
For the spectral analysis of the jets in the ACIS data,
we used the XSPEC package, v.11.2.01.
In all spectral fits we used the hydrogen column density
fixed at
$n_{\rm H}=3.2 \times 10^{20}$ cm$^{-2}$,
as determined from the
observation of the Vela pulsar with the \chan\ Low-Energy Transmission
Grating Spectrometer
(Pavlov et al.\ 2001b).
Furthermore, because of a build up of a
contaminant on either the
ACIS CCDs or the filters, additional time-dependent absorption is accounted for
in the spectral fits using the {\tt ACIS\_ABS} model
(see Plucinsky et al.\ 2002).  The degree of absorption
was set by the number of days since \chan\ launch.
Only photons in the energy range of 1--8 keV were used for fitting
to avoid contaminations from readout strips from the piled-up pulsar
and to reduce the charged particle background.

\begin{figure}
\begin{center}
\includegraphics[width=3.4in]{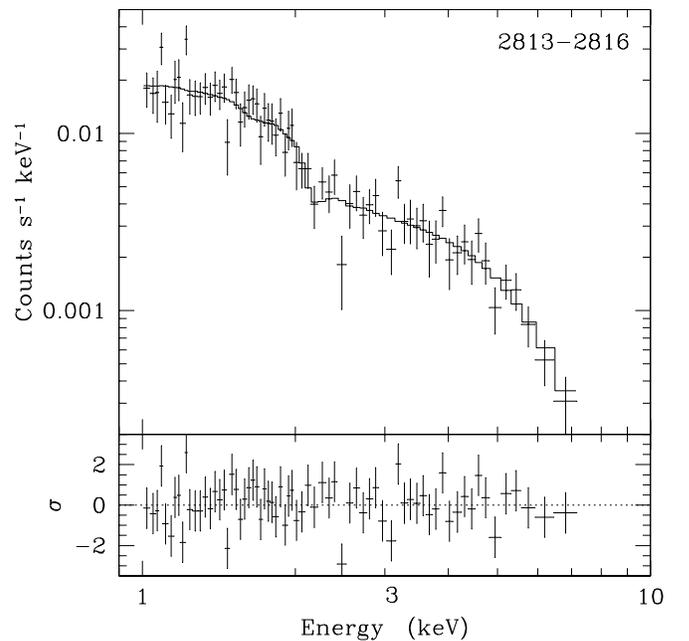}
\figcaption{Power-law fit to the spectrum of the outer jet for the
merged data set (observations 2813--2816). The best-fit parameters
are $\Gamma = 1.29\pm0.06$, and $N_\Gamma = 3.90\pm0.40 \times
10^{-5 }$ photons cm$^{-2}$ s$^{-1}$keV$^{-1}$ at 1 keV
($\chi_\nu^2= 0.98$, for 68 dof). \label{spec}}
\end{center}
\end{figure}

\begin{figure*}
\begin{center}
\includegraphics[height=4in]{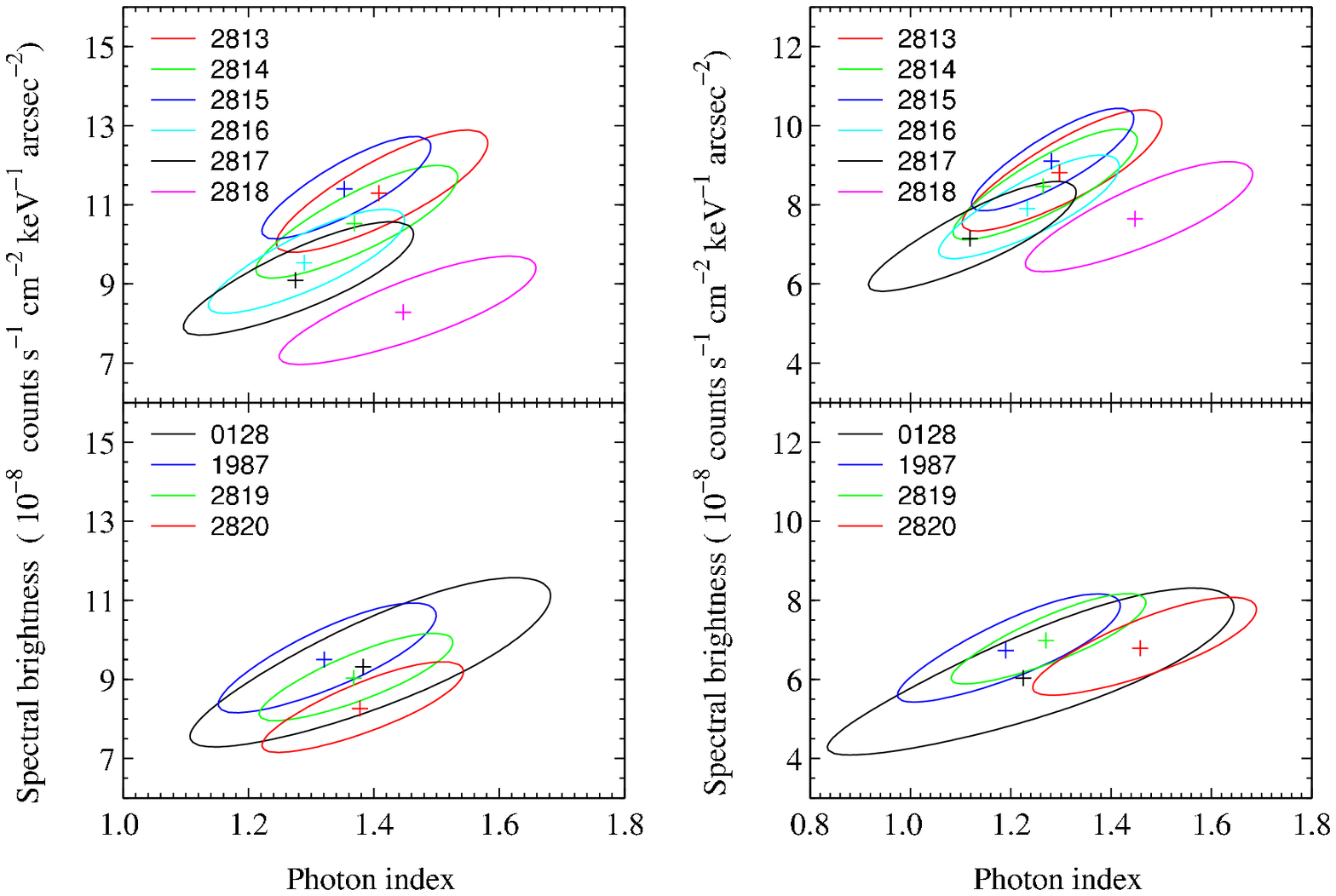}
\figcaption{One-sigma confidence contours for photon index
$\Gamma$ and spectral surface brightness at 1 keV, ${\cal
B}=N_\Gamma/A$, for 10 ACIS observations of the outer jet using
background A (left panels) and background B (right panels). The
contours are separated into the upper and lower panels for clarity
of presentation. \label{contours}}
\end{center}
\end{figure*}

The photons from the outer jet
were extracted using polygons
 (of somewhat different areas for the
different observations --- see Table 2) which
enclosed the entire structures beyond the PWN shell.
Two background regions were chosen --- the first
from a dark
region north
of the outer jet (background A), and the second sandwiching the
outer jet (background B).
Various spectral models for emission from an optically thin thermal plasma
were attempted to test if the emission
associated with the outer jet is due to an interaction of the jet
 with the ambient (SNR) gas.
The overall fits are formally acceptable, but the corresponding
plasma temperatures, about 30 keV, are very high,
in accordance with the lack of spectral lines in the spectra.
At such high temperatures
these models are equivalent to the thermal bremsstrahlung model,
which, with the statistics available,
is indistinguishable from a power-law at energies of interest.
The power-law model,
$f(E)=N_\Gamma\, (E/1\, {\rm keV})^{-\Gamma}$,
 is more physical than
the high-temperature bremsstrahlung
because it naturally describes non-thermal
(e.g., synchrotron) radiation;
therefore, we use it for the spectral analysis of the jets.

An example of the power-law fit to a data set combined of
observations 2813--2816, chosen
because of the similar extraction areas and locations on the chip, is shown
in Figure 8, for background B.
The results of the fits of the outer jet spectra with the power-law model
 for each of the individual ACIS
observations are given in Table 2, for both backgrounds A and B.
Figure 9 shows the contours of the photon index $\Gamma$ versus
spectral brightness
at 1 keV, ${\cal B} = N_\Gamma/A$ (where $A$ is the area of the extraction
region --- see Table 2). For each of the two backgrounds, the variability
of the photon index is statistically insignificant, contrary to the variability
of the brightness. Joint spectral fits (fitting all data simultaneously with
a common index and individual
normalizations) give
$\Gamma=1.36\pm0.04$ and $1.27\pm0.04$ for backgrounds A and B, respectively,
coinciding with the weighted average indices that are calculated
from the individual fits for each of the two backgrounds.
The difference between the average indices can be ascribed to systematic
effects associated with the choice of background. Since it is not immediately
clear which of the two backgrounds is more adequate, we adopt a conservative
estimate, $\Gamma = 1.3\pm 0.1$, consistent with both background choices.

\begin{table*}
\caption{Spectral parameters and surface brightnesses for the ACIS
observations of the the outer jet. \label{outer}}
\begin{center}
\begin{tabular}{lcc}
\tableline\tableline
& Background A & Background B \\
\tableline
\begin{tabular}{lc}
ObsID & Area \\
\tableline
0128 & 500 \\
1987 & 400 \\
2813 & 449 \\
2814 & 449 \\
2815 & 413 \\
2816 & 455 \\
2817 & 393 \\
2818 & 480 \\
2819 & 656 \\
2820 & 614 \\
\end{tabular}
&
\begin{tabular}{ccccc}
$\Gamma$ & $N_\Gamma$ & $S_B$ & $\chi^2_\nu$ & dof \\
\tableline
1.38$\pm$0.20 & 4.7$\pm$0.7 & 4.4$\pm$0.4 & 0.71 & 13 \\
1.32$\pm$0.12 & 3.8$\pm$0.4 & 5.7$\pm$0.3 & 1.38 & 26 \\
1.41$\pm$0.11 & 5.1$\pm$0.5 & 6.7$\pm$0.3 & 0.91 & 28 \\
1.37$\pm$0.11 & 4.7$\pm$0.4 & 7.0$\pm$0.3 & 0.79 & 31 \\
1.35$\pm$0.09 & 4.7$\pm$0.3 & 6.5$\pm$0.3 & 0.82 & 41 \\
1.29$\pm$0.10 & 4.3$\pm$0.4 & 7.3$\pm$0.4 & 0.94 & 30 \\
1.28$\pm$0.13 & 3.6$\pm$0.4 & 5.0$\pm$0.3 & 1.02 & 27 \\
1.45$\pm$0.13 & 4.0$\pm$0.4 & 3.6$\pm$0.3 & 1.01 & 26 \\
1.37$\pm$0.10 & 5.9$\pm$0.5 & 4.7$\pm$0.2 & 0.87 & 38 \\
1.38$\pm$0.10 & 5.1$\pm$0.5 & 5.0$\pm$0.3 & 0.92 & 27 \\
\end{tabular}
&
\begin{tabular}{ccccc}
$\Gamma$ & $N_\Gamma$ & $S_B$ & $\chi^2_\nu$ & dof \\
\tableline
1.23$\pm$0.27 & 3.0$\pm$0.7 & 3.2$\pm$0.4 & 0.62 & 13 \\
1.19$\pm$0.15 & 2.7$\pm$0.4 & 5.1$\pm$0.4 & 1.44 & 26 \\
1.30$\pm$0.13 & 4.0$\pm$0.5 & 5.6$\pm$0.3 & 1.01 & 28 \\
1.27$\pm$0.13 & 3.8$\pm$0.4 & 5.6$\pm$0.3 & 0.81 & 31 \\
1.28$\pm$0.10 & 3.8$\pm$0.4 & 6.0$\pm$0.3 & 0.87 & 41 \\
1.23$\pm$0.11 & 3.6$\pm$0.4 & 5.5$\pm$0.4 & 1.02 & 30 \\
1.12$\pm$0.14 & 2.8$\pm$0.4 & 5.5$\pm$0.3 & 1.04 & 27 \\
1.45$\pm$0.15 & 3.7$\pm$0.4 & 4.5$\pm$0.3 & 1.02 & 26 \\
1.27$\pm$0.13 & 4.6$\pm$0.5 & 4.5$\pm$0.3 & 0.85 & 38 \\
1.46$\pm$0.15 & 4.1$\pm$0.5 & 3.9$\pm$0.3 & 1.05 & 27 \\
\end{tabular}
\\
\tableline
\end{tabular}
\end{center}
\begin{small}
Area is the polygon extraction area of the outer jet in
arcsec$^2$. Normalization of the power-law spectrum, $N_\Gamma$,
is in units  of $10^{-5}$ photons cm$^{-2}$ s$^{-1}$ keV$^{-1}$ at
1 keV. Mean surface brightness, $S_B$, is in units of 10$^{-2}$
counts arcsec$^{-2}$ ks$^{-1}$, in the 1--8 keV range.
\end{small}
\end{table*}

We searched for a change in power-law index along the length of the outer jet
by extracting regions near the beginning of the outer jet, where it exits the
shell of the bright PWN, and at its end.  Because of the low number of counts, a joint
fit of all the ACIS observations
was attempted.  The normalizations for each observation were allowed to
vary (because the extraction areas were different),
while the photon index for all
observations was fit to a common value.  The fits to these combined spectra
were statistically acceptable, with $\chi_\nu^2$=1.19 (82 degrees of
freedom [dof]) and $\chi_\nu^2=0.97$ (96 dof)
for the beginning
and the end of the outer jet, respectively (using background B).
The change of the index, from $1.36\pm0.09$ at the jet
beginning to $1.25\pm0.08$ near its end, is not statistically
significant.

We also attempted to compare the spectra of the blobs with each other
and with the spectrum of the rest of the outer jet. For individual
observations, no differences are seen because of the low number of counts.
Combined fits, using background B,
yield the photon indices $1.36\pm0.12$ ($\chi_\nu^2$=1.01, 70 dof) for blob A
(observations 2813--2818), $1.07\pm0.19$ ($\chi_\nu^2$=0.47, 20 dof)
for blob B (observations 2813--2816), and $1.43\pm0.21$ ($\chi_\nu^2$=1.27, 18 dof)
for blob C (observations 2813--2815).
Thus, the spectral slopes do not show statistically significant differences
even for the combined fits.

 For the inner jet and inner counter-jet, we extracted the source
counts from stretched elliptical areas of 12.7 and 18.5 arcsec$^2$,
respectively, which
enclose as much of the jet/counter-jet emission
 as possible without contamination from
the other structures within the PWN, such as the arcs.
Because of the very nonuniform distribution of surface brightness
around the inner jet and counter-jet, we had to take
backgrounds from small regions; we chose circles (6.8 and 10.6 arcsec$^2$
for the inner jet and inner counter-jet, respectively) in the immediate
vicinities northeast and southwest of the source extraction
regions.
The average values of the photon index for the inner jet
and inner counter-jet, $\Gamma= 1.09$ and 1.20, respectively,
appear somewhat smaller than that for the outer jet. However, the photon indices
measured in individual observations show large deviations from the average
values (rms deviations are 0.24 and 0.10, respectively),
likely associated with statistical fluctuations of the bright background
estimated from small regions.
Therefore, we cannot firmly establish the spectral differences
between the outer jet and its inner counterparts from the data available.

As the outer counter-jet is hardly seen in the individual images,
its spectrum was measured in the joint fit
using observations 2813--2820.
The source counts were extracted from
a polygon area of 336 arcsec$^2$.
Because of the faintness of the outer counter-jet, the fit is rather
sensitive to the choice of background. Taking background from a region
northeast of the end of the outer counter-jet, we obtain
the photon index
$\Gamma\approx 1.5$, while the index is smaller,
$\Gamma\approx 1.2$, for the
background from a region
which sandwiches the outer counter-jet.
Thus, the spectrum of the outer counter-jet remains poorly constrained,
its slope being indistinguishable from that of the outer jet, within the
uncertainties.

Assuming isotropic radiation, the average (unabsorbed) luminosities of the
inner jet, inner counter-jet, and outer counter-jet,
 in the 1--8 keV band,
are 2.6, 4.5 and 0.5 $\times 10^{30}$ erg s$^{-1}$,
respectively, for $d=300$ pc.
The average luminosities are computed
from the average spectral parameters of the various components.
The outer jet
demonstrates some variability in luminosity.  For background A,
the extremes of the luminosity are $2.7\pm0.3$ and $4.4\pm0.4$
$\times 10^{30}$ erg s$^{-1}$, with an
average of $3.4\times 10^{30}$ erg s$^{-1}$.
For background B, the outer jet luminosity
 varies from $2.4\pm0.3$ to $3.8\pm0.4$ $\times 10^{30}$ erg s$^{-1}$,
 with an average of
$3.0\times 10^{30}$ erg s$^{-1}$.
 For comparison, the luminosity of the whole PWN within $42''$ of the pulsar
is $L_{\rm pwn}=6.0\times
10^{32}$ erg s$^{-1}$, in the same energy band.

\section{ Discussion.}
The detected variability of the outer jet is the most vivid
demonstration of the dynamical behavior of the Vela PWN.
The observed changes of the overall shape of the outer jet
over a time scale of weeks
 suggest apparent speeds up to (0.3--0.7)$ d_{300}\, c$,
which is comparable to the apparent speeds of the blobs moving along
the outer jet away from the pulsar.
The spectra of the outer jet, as well as those of the other PWN elements,
fit well with a power-law model, indicating that this is radiation from
ultrarelativistic particles.
These observations allow one to assume that the outer jet is associated
with a polar outflow of relativistic particles from the pulsar's
magnetosphere\footnote{An alternative interpretation --that the ``outer jet''
is a limb-brightened shell (northeast boundary) of a large diffuse nebula
southwest of the pulsar (see Fig.\ 1) or a shock front-- looks very unlikely,
given the observed variations of its shape and, particularly, the moving blobs.}.
Possible implications of our observational results are discussed below.
\subsection{ Geometry }
The orientation of the jets
is the most important item for
explaining the observed dynamical behavior, evaluating the velocities,
and calculating the energetics of these outflows.
Immediately after the first {\sl Chandra} observations of the
Vela PWN, it was suggested that the bright (inner) jet and counter-jet are co-aligned
with the rotational axis of the pulsar
and, presumably, with the direction of pulsar's velocity (Pavlov et al.\ 2000;
Helfand et al.\ 2001).
 As we have seen from Figures 3 and 4,
the projection of the outer jet on the sky plane may strongly deviate from the
(straight) extension of the
projection of the inner jet/counter-jet. However,
the outer jet projection was close to that line during the period of
2001 November 25 through 2002 January 13, when
the apparent velocities of the bright blobs were measured.
Assuming, by analogy with the jets in AGNs and Galactic microquasars,
that the apparent motion of blobs is associated with
motion of matter along
an approximately straight line
(e.g., the three-dimensional velocities of the blobs coincide with the
bulk flow velocity),
one can constrain
the true speed,
$v=\beta c$,
and the angle $\theta$ between the line of sight and the
direction of motion. The apparent speed, $v_{\rm a}=\beta_{\rm a} c$, is related
to the true speed
by the equation $\beta_{\rm a} = \beta\, \sin\theta\,
(1-\beta\, \cos\theta)^{-1}$. Figure 10 shows the dependence $\beta(\theta)$
for typical $\beta_{\rm a}$
inferred from the observed
motions of the blobs.
For $0.3 < \beta_{\rm a} <0.7$,
the blobs can be approaching ($\theta<90^\circ$) as well as receding
($\theta>90^\circ$). The range
of minimum values of $\beta$ is $0.28 < \beta_{\rm min} < 0.57$
(at $73^\circ > \theta > 55^\circ$), while the
range of maximum allowed $\theta$ (at which $\beta \to 1$),
 is $147^\circ > \theta_{\rm max} > 110^\circ$. The latter values are smaller
than the $\theta = 155^\circ$, inferred by Helfand et al.\ (2001)
under the assumption that the PWN arcs are Dopler-brightened parts
of a torus around the jet/counter-jet, and they are smaller then
the $\theta = 152^\circ$ suggested by Pavlov (2000)\footnote{see
also {\em
http://online.itp.ucsb.edu/online/neustars\_c00/pavlov/oh/34.html}}
 assuming the arcs are Doppler-brightened
parts of ring-like shocks in relativistic conical outflows. If we adopt,
e.g., $\theta=155^\circ$, then the maximum possible apparent speed
(at $\beta \to 1$) is $0.22\, c$,
which is below the lower limits
of $0.30\,c$ and $0.38\,c$ on the
apparent speeds of blobs A and B at $d=300$ pc.
Thus, if any of the above-mentioned interpretaions of the
arcs and
the inner jet/counter-jet are correct,
we have to conclude that the outer jet is tilted
from the pulsar rotation axis towards the observer
even if its projection onto the sky looks almost
straight, like in panels 5--9 of Figure 4. Moreover, we cannot
rule out the possibility that the outer jet
in these observations is strongly bent
in the plane perpendicular to the sky plane --- e.g., similar to the jet's
sky projections in panels 1, 2 or 13. In this case the directions of
the local flow velocities vary along the jet,
which might explain the different apparent
velocities of the blobs A and B.
Such bending opens up a possibility that
the bright regions of the jet are
merely due to the projection effect --- the segments of a uniformly bright, bent
jet, which are oriented along the line of sight, appear
brighter because of the increased
optical depth, and the brightness is further enhanced by the Doppler boosting
if the flow in these segments is streaming toward the observer.
 For instance, if the loop-like jet
in panel 2 of Figure 3 were a two-dimensional stucture,
the southern segment of the loop would look like a bright blob
if it were observed from the southwest direction.
If this interpretation is correct, then the observed motions are caused
by changing geometrical shape of the jet rather than the flow of
locally bright material along the jet.
An argument against such an interpretation is that the outer jet
apparently terminates
with a bright blob in at least 10 of 13 observations, and it is hard to
believe that the end segment of a randomly bent jet is
so often directed toward the observer.  However, the extension of the jet beyond
these bright blobs, as seen in the summed image of Figure 1
(see also \S3.3), suggests that this may
still be the case.

\begin{figure}
\begin{center}
\includegraphics[width=3.4in]{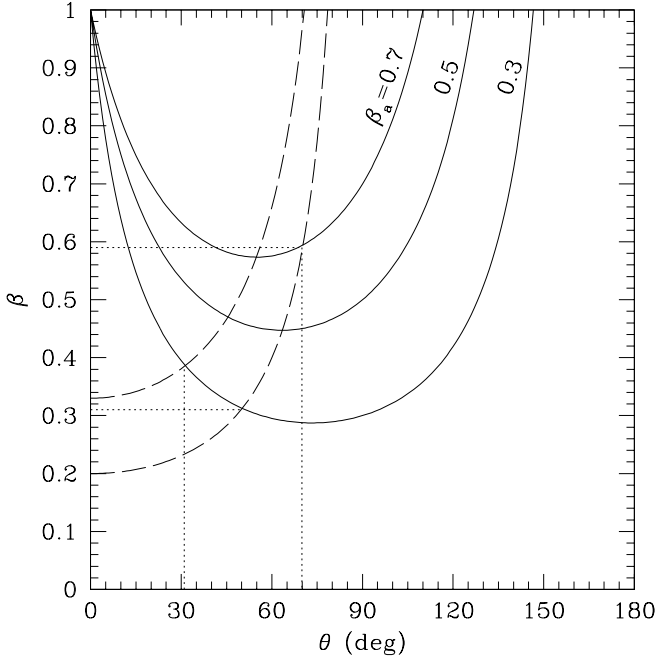}
\figcaption{True speed $\beta=v/c$ versus angle $\theta$ between
the line of sight and direction of motion for three values of the
apparent speed $\beta_{\rm a}=v_a/c$. The dashed lines ($\beta
\cos\theta =0.20$ and 0.33) bracket the domain of allowed
$\beta$,$\theta$ assuming intrinsically similar outer jet and
outer coun ter-jet streaming in opposite directions. See text for
details. \label{beta} }
\end{center}
\end{figure}

\begin{figure}
\begin{center}
\includegraphics[width=3.4in]{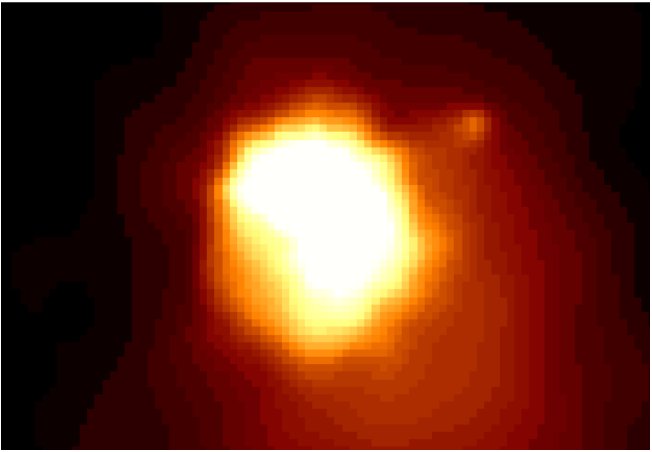}
\figcaption{Smoothed {\sl ROSAT} HRI image ($5\farcm7\times
3\farcm8$) of the Vela PWN from the observation of November of
1997 (exposure 33 ks). The blob to the northwest of the PWN is
indication that the outer jet has persisted at least 5 years.
\label{rosat}}
\end{center}
\end{figure}

Another interesting fact related to the jet geometry is that
the outer jet is consistently brighter
than the outer counter-jet, with the ratio of surface brightnesses
$f_{\rm b}=4$--9
(the large uncertainty in the brightness ratio is due to the nonuniform
backgrounds).
If the two jets are energetically similar and, on average, are oriented
along a straight line, then the outer jet is approaching
while the outer counter-jet is receding, contrary to the receding inner jet and
approaching inner counter-jet in the geometry assumed by Pavlov et al.\ (2000)
and Helfand et al.\ (2001).
Within this interpretation, the brightness ratio is
$f_{\rm b}=[(1+\beta\,\cos\theta)/(1-\beta\,\cos\theta)]^{\Gamma +2}$,
where $\Gamma$ is the photon index.
 For $f_{\rm b}=4$--9 and $\Gamma=1.2$--1.4, it gives
$\beta\,\cos\theta = 0.20$--0.33,
$\theta < 71^\circ$--$78^\circ$.
Furthermore, these values of $\beta\cos\theta$ are consistent with
 $\beta_{\rm a} = 0.3$--0.7, inferred from the blob's motion, if
$0.31 < \beta < 0.59$ and $31^\circ < \theta < 70^\circ$ (see Fig.\ 10).
To reconcile this result with the previously
suggested orientation of the inner jets,
one has to assume that the outer jet is tilted by a large angle
($\approx 90^\circ$--$100^\circ$) from the inner jet direction towards the observer,
while the outer counter-jet is tilted by a similar angle away
from the observer.
Contrary to the observed variable bendings of the outer jet, such tilts
would have persisted for at least 5 years, since we confirm the existence
of the brighter outer jet as far back as the 1997 {\sl ROSAT} HRI observation
(Fig.\ 11). It is not clear what could cause such persistent tilts.
Furthermore, it not clear why the outer jet and
counter-jet are so much dimmer than their inner counterparts, and
what causes the abrupt decrease of brightness in the
outer jet and counter-jet, with bright knots
at the junctions of the inner and outer parts.
In particular, the transition
from the inner  jet
to the outer jet apparently occurs at the intersection
of the jet with the outer arc, which hints that the outer arc is not
a part of a ring-like structure but an umbrella-like shell pierced by the jet
(Kargaltsev et al.\ 2002).
Such a shell could be a bow shock created by the (inner) jet in the
ambient medium, or it could form as a result of compression of the plasma outflow
(hence, amplification of the frozen-in magnetic field) by the
pulsar's motion in the ambient SNR matter (similar to that suggested by Aschenbach
\& Brinkmann 1975 for the Crab PWN).
Another possible effect of the pulsar's motion is distortion of the ring-like
structure(s) (e.g., the post-shock region presumably associated with the inner
arc) by the ram pressure of the ambient matter, if the pulsar's rotational axis
is not co-aligned with its velocity. In this case the leading part of the ring
can be flattened by the ram pressure so that the assumption that the
ring's sky projection is a perfect ellipse can result in wrong values for the inclination
angles. Finally, one could consider a possibility that the structures appearing
as the bright  inner jet and
counter-jet are, in fact, merely traces of particle beams
in conical outflows brightened by the
Doppler boosting (e.g., Radhakrishnan \& Deshpande 2001; Pavlov 2000), while
the outer jet and counter-jet are the ``true pulsar jets''. Such an
assumption implies a quite different interpretation of the whole PWN.
We will discuss these and other possible PWN models in a future paper.
Currently, we cannot unambiguously determine the orientation of the jets or
the true velocities. However, from the observed variations of the shape
of the outer jet, with mildly relativistic apparent velocities, we can
conclude that the true velocities are neither ultrarelativistic nor nonrelativistic;
$0.3\,c$ and $0.7\,c$ can be adopted as conservative lower and upper
limits.

\subsection{
Magnetic field and energetics}
The power-law spectra of the outer jet
and the other PWN elements can be interpreted as
optically thin synchrotron emission from ultrarelativistic electrons (and/or positrons)
with a power-law spectrum: ${\rm d}N(\gamma) =K \gamma^{-p}\,{\rm d}\gamma$,
where $\gamma=\varepsilon/m_ec^2$ is the Lorentz factor,
$\gamma_{\rm m} < \gamma < \gamma_{\rm M}$.
The electron index $p$ is related to the photon index
as $p=2\Gamma-1$, i.e., $p=1.4$--1.8 for $\Gamma\simeq 1.2$--1.4
observed in the outer jet.
A characteristic energy $E$ of the synchrotron photon depends on $\gamma$ and
magnetic field $B$
as $E\sim 5\, B_{-4} \gamma_8^2$ keV, where
$B_{-4}=B/(10^{-4}\,{\rm G})$, $\gamma_8=\gamma/10^8$.
If the minimum and maximum energies of the photon power-law spectrum
are $E_{\rm m}$ and $E_{\rm M}$, respectively, then the corresponding
boundaries of the electron power-law spectrum are
$\gamma_{\rm m} \approx 2.4\times 10^7\, [E_{\rm m}/(y_{{\rm m},p}\delta)]^{1/2} B_{-4}^{-1/2}$
and
$\gamma_{\rm M} \approx 2.4\times 10^7\, [(E_{\rm M}/(y_{{\rm M},p}\delta)]^{1/2} B_{-4}^{-1/2}$,
where the photon energies are in units of keV,
$\delta =
(1-\beta^2)^{1/2}(1-\beta\,\cos\theta)^{-1}$
is the Doppler factor ($\beta\,c$ is the speed of the bulk motion),  and
$y_{{\rm m},p}$ and $y_{{\rm M},p}$ are dimensionless factors.
The values of these factors are given in Table II of
Ginzburg \& Syrovatskii (1965; GS65 hereafter); e.g.,
$y_{{\rm m},p}=1.3$ and 1.8, $y_{{\rm M},p} = 0.011$ and 0.032,
for $p=1.5$ and 2.0, respectively.
We will use the well-known formulae for synchrotron radiation to estimate
the magnetic field and energetics of the outer jet. Since the bulk motions observed
in the jet are only mildly relativistic ($\beta\sim 0.5$ is a plausible
estimate), the Doppler factor
is not strongly
different from unity;
moreover, its actual
value is unknown because of the uncertain orientation of the jet.
Therefore, we will neglect the bulk motion of the jet's matter for
most of the estimates below.

 From the {\sl Chandra} ACIS observations,
the power-law spectrum of the outer jet is seen in the energy range
between $E_1\approx 0.5$ keV and $E_2\approx 8$ keV.
Since $E_{\rm m} < E_1$ and $E_{\rm M} > E_2$,
we can constrain the minimum and maximum energies of the electron power-law
distribution:  $\gamma_{\rm m} < 1.4\times 10^7\, B_{-4}^{-1/2}$,
$\gamma_{\rm M} > 5\times 10^8\, B_{-4}^{-1/2}$.

One can put a lower limit on the magnetic field in the outer jet
using the condition that the Larmor radius of most energetic
electrons responsible for the X-ray emission,
$r_L = 1.7\times 10^{15} \gamma_8 B_{-4}^{-1}\,\,{\rm cm}
\ga 6\times 10^{15} (E_{2}/8\,{\rm keV})^{1/2} B_{-4}^{-3/2}$ cm, does
not exceed the jet radius, $r_{\rm jet}\approx 1.5\times 10^{16} d_{300}$ cm.
This condition gives $B \ga 60\, (E_{2}/8\,{\rm keV})^{1/3} d_{300}^{-2/3}$ $\mu$G.

An upper limit on the magnetic field can be estimated from the fact
that the spectrum of the outer jet maintains its power-law shape
(shows no appreciable spectral break)
at the observed energies $E\la 8$ keV up to the end of the outer jet.
 This means that the time of synchrotron losses,
$t_{\rm syn}=5.1\times 10^8 \gamma_8^{-1} B_{-4}^{-2}$ s, for the
most energetic
particles responsible for the observed X-ray spectrum [$\gamma_8 \ga
4\, B_{-4}^{-1/2}(E_2/8\,{\rm keV})^{1/2}$]
is longer (or comparable to)
the flow time, $t_{\rm flow}=l_{\rm jet}/v_{\rm flow} \approx 6.7\times 10^7
l_{\rm jet,18} (\beta/0.5)^{-1}$ s, where $l_{\rm jet}=10^{18} l_{\rm jet,18}$ cm.
 This gives
$B\la 150\, (E_2/8\,{\rm keV})^{-1/3}
(\beta/0.5)^{2/3} l_{\rm jet,18}^{-2/3}$ $\mu$G, i.e., $B \la 100$--300 $\mu$G
for plausible values of $l_{\rm jet}$ and $\beta$.

An independent estimate on the plasma parameters for the outer jet
can be obtained
from the brightness of the synchrotron radiation,
assuming some value for the ratio $k_m=w_{\rm mag}/w_{\rm rel}$ of the
magneic energy density,
$w_{\rm mag}= B^2/(8\pi)$, to the energy density of relativistic particles,
$w_{\rm rel}$.
We cannot exclude the possibility that there are some relativistic
ions
in the jet, in addition to the synchrotron-emitting electrons
(e.g., Gallant \& Arons 1994), so that
$w_{\rm rel} = (1+k_i) w_e$, where $k_i$ is the ratio
of the energy density of ions to the energy density of
electrons, $w_e$.
Making use of the standard formulae for synchrotron radiation
(e.g., GS65;
Pacholczyk 1970), one
can express the magnetic field in terms of the observable parameters.
For a tangled magnetic field (randomly distributed along the line of sight),
we obtain
\begin{equation}
B=27\,\left\{\frac{k_m (1+k_i)}{a_p (3-2\Gamma)}
\left[E_{{\rm M},p}^{(3-2\Gamma)/2} - E_{{\rm m},p}^{(3-2\Gamma)/2}\right]
\frac{\mathcal{B}_{-7}}{\bar{s}_{16}}\right\}^{2/7}\,\,\mu{\rm G}\,,
\end{equation}
where
$\mathcal{B}=N_\Gamma/A = 10^{-7} \mathcal{B}_{-7}$ photons
(s\,cm$^2$\,keV\,arcsec$^2$)$^{-1}$ is the average spectral surface
brightness
at $E=1$ keV,
$N_\Gamma$ is the normalization of
the photon spectral flux measured in area $A$,
$\bar{s}=10^{16}\, \bar{s}_{16}$ cm is an average length of the radiating
region along the line of sight, $E_{{\rm m},p}=E_{\rm m}/y_{{\rm m},p}$,
$E_{{\rm M},p}=E_{\rm M}/y_{{\rm M},p}$, $E_{\rm m}$ and $E_{\rm M}$ are
the lower and upper energies of the photon power-law spectrum (in keV),
and $a_p$ is a numerical coefficient given by eq.\ (3.32) and Table II of GS65 (e.g.,
$a_p =0.165$ and 0.117 for $p=1.4$ and 1.8 [$\Gamma=1.2$ and 1.4], respectively).
The value of $B$ depends on several unknown parameters. Particularly
uncertain are the
boundary energies of the photon power-law spectrum because of the lack of
high-resolution observations outside the X-ray range.
Fortunately, the dependence on these energies is rather weak:
$B\propto E_{\rm M}^{(3-2\Gamma)/7}$
for $E_{{\rm m},p}^{1.5-\Gamma} \ll E_{{\rm M},p}^{1.5-\Gamma}$
(which implies $\Gamma < 1.5$), and
$B\propto \left[\ln (56 E_{\rm M}/E_{\rm m})\right]^{2/7}$,
for $\Gamma\simeq 1.5$. For $\mathcal{B}_{-7}\simeq 1$
(a typical spectral surface brightness at 1 keV --- see Fig.\ 9),
 $\bar{s}_{16}\approx 2$--10 (the main
reason of the uncertainty is the unknown spatial structure and orientation
of the jet), and $\Gamma = 1.2$--1.4, we obtain
$30\,\mu{\rm G} \la B[k_m (1+k_i)]^{-2/7} \la 200\,\mu{\rm G}$,
for plausible ranges of $E_{\rm M}$ and $E_{\rm m}$.
It is usually assumed that $k_m\approx 1$ (equipartition condition),
while the value of $k_i$ is rather uncertain. For the lower limit on the equipartition
field not to exceed the upper limit estimated from (the lack of) synchrotron
cooling, $k_i$ should not exceed  $\sim 10^3$. Thus, given the uncertainty
of the parameters, we can only state that the equipartion field is consistent
with the above-estimated limits, and that a plausible estimate for a typical
field in the jet is $B\sim 100\,\mu$G, with an uncertainty of a factor of 3.
It should be noted that the local field values can differ from the average
field. For instance, if the brightness of the blobs is due to an increased
magnetic field, at fixed $\bar{s}$, then the field in the blobs is a factor
of 1.4 higher than the average field in the jet. On the other hand, the higher
brightness can be explained by the projection effect (larger $\bar{s}$) and
the Doppler boost ($\bar{s} \to \bar{s}\, \delta^{2+\Gamma}$ in eq.\ [1] if the
bulk motion is taken into account).

Assuming equipartition,
$w_{\rm rel} = B^2/(8\pi)= 4\times 10^{-10}\, B_{-4}^2$ erg cm$^{-3}$,
we can estimate the total energy of relativistic particles,
$W_{\rm rel} =
4\times 10^{41}
B_{-4}^2\, V_{51}$ erg, where $V=10^{51} V_{51}$ cm$^3$
is the volume of the jet. For plausible values of
the magnetic field and the volume ($V_{51} \sim 0.2$--1,
depending on the shape and orientation of the jet),
we obtain $W_{\rm rel} \sim 10^{40}$--$10^{42}$ erg.
The energy injection rate and the energy flux can be estimated as $\dot{W}
 \sim 2 W_{\rm rel}/t_{\rm flow}
\approx 8\times 10^{33}\, B_{-4}^2 (r_{\rm jet}/1.5\times 10^{16}\, {\rm cm})^2 (\beta/0.5)$
erg s$^{-1}$ and $F_W = \dot{W}/(\pi r_{\rm jet}^2) \approx 10\, B_{-4}^2 (\beta/0.5)$
erg cm$^{-2}$ s$^{-1}$, respectively, where $r_{\rm jet}$
is the jet radius, and $\beta c$ is the
flow speed. This injection rate is a small fraction of the spin-down
energy loss rate of the Vela pulsar, $\dot{W} \sim 10^{-3} \dot{E}$.
On the other hand, it greatly exceeds the observed X-ray luminosity of the jet,
$L_{\rm x,jet} \sim 10^{-3} \dot{W}$, which means that most of
the injected energy is emitted outside the X-ray range or, more probably,
carried out of the jet without substantial radiation losses.

One can also estimate the electron (positron) injection rate,
$\dot{N}_{e,{\rm jet}} = \dot{W}[2 (1+k_i) m_ec^2 \bar{\gamma}]^{-1}$,
where
$\bar{\gamma}= [(p-1)(\gamma_M^{2-p}-\gamma_m^{2-p})]
[(2-p)[(\gamma_m^{1-p} - \gamma_M^{1-p})]^{-1}$
is the average electron Lorentz factor for the power-law distribution.
For the mean observed $\Gamma\simeq 1.3$, we have
$\bar{\gamma}\approx 1.5\, \gamma_M^{0.4} \gamma_m^{0.6}$ if
$\gamma_m^{0.4} \ll \gamma_M^{0.4}$.
Unfortunately, the minimum and maximum electron energies cannot be
determined without multiwavelength observations of the jet, therefore
we will scale
$\bar{\gamma}$ to a possible (but arbitrary)
value of $10^8$, which gives $\dot{N}_{e,{\rm jet}} \approx 5\times 10^{31}\,
\bar{\gamma}_8^{-1} (1+k_i)^{-1}$ s$^{-1}$. This estimate corresponds to
the mean electron number density $n_{e,{\rm jet}} \sim
w_{\rm mag} (1+k_i)^{-1} (m_e c^2 \bar{\gamma})^{-1}\sim
5\times 10^{-12} B_{-4}^2 (1+k_i)^{-1} \bar{\gamma}_8^{-1}$ cm$^{-3}$.
The numerical estimates for $\dot{N}_{e,{\rm jet}}$ and $n_{e,{\rm jet}}$
strongly depend on the boundary energies of the electron power-law
spectrum; in particular, the estimates become 3 orders of magnitude
larger if $\gamma_{\rm m}$ is low enough ($\sim 10^3$) for the jet to
be synchrotron-emitting in the radio band.

It is interesting to compare the estimate for $\dot{N}_{e,{\rm jet}}$
 with the pair production rate
expected for the Vela pulsar: $\dot{N}_{e,\rm puls}
\sim  n_{\rm GJ} (4\pi^2 R^3/P)\kappa_{\rm pair} =
4\pi^2 R^3 B_{\rm puls}(ceP^2)^{-1} \kappa_{\rm pair}
\simeq 1.2\times 10^{33}\, \kappa_{\rm pair}$ s$^{-1}$,
where $n_{\rm GJ}$ is the Goldreich-Julian density, $R\approx 10^6$ cm and
$P= 0.089$ s are the neutron star radius and spin period,
 $B_{\rm puls}=3.4\times 10^{12}$ G
is the pulsar magnetic field, and $\kappa_{\rm pair}$ is the pair multiplication
coefficient (its plausible value is $\kappa_{\rm pair} \sim 10^3$  --- see, e.g.,
Hibschman \& Arons 2001). The ratio $\dot{N}_{e,{\rm jet}}/\dot{N}_{e,\rm puls}
\sim 4\times 10^{-5}\, (\kappa_{\rm pair}/10^3) \bar{\gamma}_8^{-1} (1+k_i)^{-1}$
can be interpreted as the fraction of pairs escaping from the pulsar through the
outer jet, assuming that no pairs are created in the jet itself. It is worth
noting that the maximum Lorentz factor for the pairs created in the pulsar
magnetosphere does not exceed a few $\times 10^7$, for both the polar cap models
(Harding, Muslimov, \& Zhang 2002, and references therein) and the outer gap
models (e.g., Zhang \& Cheng 1997). The maximum Lorentz factor in the outer jet,
$\gamma_{\rm M} > 5\times 10^8$, is substantially higher, which means that
the pairs, created by the pulsar, have been additionally accelerated beyond
the pulsar magnetosphere.

If ions are pulled out of the neutron star surface layers, their outflow rate
is constrained to the Goldreich-Julian value: $\dot{N}_{i,\rm puls} \sim
\dot{N}_{e,\rm puls}(Z \kappa_{\rm pair})^{-1}
\simeq 1.2\times 10^{33}\, Z^{-1}$ s$^{-1}$,
while their characteristic energy may exceed that of electrons/positrons
(Gallant \& Arons 1994). Although the ions can be further accelerated to
even higher energies, the maximum ion energy in the jet cannot exceed
$\varepsilon_{i, \rm M} \sim ZeB r_{\rm jet} \sim 5\times 10^{14}\,Z B_{-4}
(r_{\rm jet}/1.5\times 10^{16}\, {\rm cm})$ eV
[$\gamma_{i, \rm M}\sim 5\times 10^5\, (Z/A) B_{-4}
(r_{\rm jet}/1.5\times 10^{16}\, {\rm cm})$] because
the Larmor radius of ions with higher
energies is larger than the jet radius. Assuming that the ion-to-electron number
density ratio in the jet does not exceed that
in the magnetosphere outflow,
we can constrain the ion-to-electron energy density ratio:
$k_i=(n_{i,\rm jet}\bar{\varepsilon}_{i,\rm jet})/(n_{e,\rm jet}
\bar{\varepsilon}_{e,\rm jet}) < (\dot{N}_{i,\rm puls}/\dot{N}_{\rm e,\rm puls})
(ZeBr_{\rm jet}/m_ec^2\bar{\gamma}) \sim  10^{-2}\,
B_{-4} \bar{\gamma}_8^{-1}
(10^3/\kappa_{\rm pair})$, at $r_{\rm jet}=1.5\times 10^{16}$ cm.
This means that relativistic ions do not make a substantial contribution to
the energetics of the outer jet.

Since the bulk velocities in the outer jet are only mildly relativistic,
the bulk kinetic energy of relativistic particles is much lower than the
energy of their random motion. However, we cannot rule out the possibility
that some nonrelativistic electron-ion plasma from the ambient SNR
medium is entrained into the jet via interaction with the
jet's magnetic field.
To estimate an upper limit on the maximum density of the nonrelativistic
component and maximum kinetic energy of the outer jet,
it seems reasonable to assume that the {\em total} energy injection
rate, including the kinetic energy, into the outer jet cannot exceed
a fraction of $L_{\rm jet}/L_{\rm pwn}\approx 0.008$ of the total power, $\dot{E}
\approx 7\times 10^{36}$ erg s$^{-1}$,
supplied by the pulsar, which implies
$W_{\rm kin} < 0.008\,\dot{E} t_{\rm flow} \sim 4\times 10^{42}\, l_{\rm jet, 18}
(\beta/0.5)^{-1}$ erg.
This condition requires densities of nonrelativistic particles (ions or electrons),
$n_{\rm nonrel} < 2\times 10^{-4}\, (r_{\rm jet}/1.5\times 10^{16}\, {\rm cm})^{-2}
(\beta/0.5)^3$ cm$^{-3}$.
Thermal X-ray radiation from such a low-density plasma is orders of magnitude
fainter than the observed radiation from the outer jet.
 The corresponding upper limits on the pressure
and energy density of the nonrelativistc components depend on the unknown
mean energy of nonrelativistic particles; however, they do not exceed
the pressure and energy density of relativistic particles and the magnetic field.
Furthermore, the density
of this plasma is much lower than the density of the ambient medium.
On the other hand,
the bulk pressure, $p_{\rm bulk} \sim \rho v_{\rm bulk}^2
< 7\times 10^{-8}$ erg cm$^{-3}$, can be higher than the pressure in the
ambient medium, $p_{\rm amb} \sim 10^{-9}$ erg cm$^{-3}$.
\subsection{
End bend and outer PWN: Effect of SNR wind?}
We see from Figures 3 and 4 that the outer jet is never straight, showing
either gentle bends (e.g., panels 5--10) or a strongly curved structure
(panels 1, 2, and 13). The varying curvature of the outer jet can be explained
by its interaction with the ambient medium (SNR plasma) and/or by a kink instability
in the jet flow.

A strong argument for
an external wind to cause jet's bending is provided by the deep image of
the PWN (Fig.\ 1) that allows one to see an extension of the outer jet
beyond the apparent termination points (often associated with outermost
blobs) observed in the individual images.
It shows that the outer jet does {\em not} terminate abruptly, but it rather
smoothly bends clockwise (towards west and then southwest) by at least
$90^\circ$. Such bending suggests a persistent {\em northeast wind} in the ambient
medium (approximately perpendicular to the direction of the pulsar's proper
motion. Such a wind could also explain the very asymmetric diffuse emission
outside the bright PWN --- the wind ``blows off'' relativistic electrons,
produced in the bright PWN, towards the southwest. Moreover, it could explain,
in the same way, the fact that the surface brightness is substantially
higher at the southwest side of the outer jet compared with the northeast ---
some high-energy particles are leaking from the jet and blown away by the
wind. Finally, an additional support for the wind comes from the radio image
of the (outer) PWN (Lewis et al.\ 2002; Dodson et al.\ 2003b).
This image shows two lobes,
northeast and southwest of the X-ray bright PWN, with
a much smaller northeast lobe confined by a brightened northeast boundary.
The smaller size and the brightening can be explained by compression of
the radio-emitting plasma by the external northeast wind. The bulk pressure
of this wind can be crudely estimated as $P_{\rm wind} \sim (d_{\rm jet}/R_{\rm curv})
(\Theta/A_\perp)$, where
$d_{\rm jet}$ is the diameter of the jet,
 $R_{\rm curv}$ is the curvature radius of the bending, and  $\Theta$ is the
thrust (jet's momentum flux) through the jet's transverse area $A_\perp$ (e.g.,
Leahy 1991).
If the bulk kinetic pressure is much lower than the magnetic pressure
within the jet, $\rho_{\rm jet} v^2 \ll B^2/(8\pi) \sim 10^{-10}$ erg cm$^{-3}$,
then the main contribution to the thrust comes from the longitudinal component
of the magnetic stress tensor that, in turn, depends on the geometry of
the magnetic field.
For instance,
we obtain $\Theta\sim A_\perp B^2/(24\pi)$ for a tangled magnetic field, which gives
$P_{\rm wind}
\sim 3\times 10^{-12} B_{-4}^2$
erg cm$^{-3}$, for $R_{\rm curv}\sim 3\times 10^{17}$ cm.
This bulk pressure
is much lower than the typical thermal pressure in the Vela SNR. It
corresponds
to the wind velocity $v_{\rm wind} \sim 10\, n_{\rm wind}^{-1/2} B_{-4}$ km s$^{-1}$,
where $n_{\rm wind}$ is the ion (proton) number density in the wind.
Attempts at separating
the wind's X-ray emission from the SNR background did not yield conclusive
results.
Assuming a low-density wind, $n_{\rm wind} \sim 10^{-2}$ cm$^{-3}$,
 we obtain
$v_{\rm wind} \sim 100\, B_{-4}$ km s$^{-1}$ --- high but not improbable velocity.
Such a velocity is comparable with the pulsar's proper motion velocity, $\simeq
97\, d_{300}$ km s$^{-1}$ (Caraveo et al.\ 2001), but the pulsar's motion itself
cannot initiate the observed bending because the pulsar moves in the direction
of the unbent jet (at least in the sky projection). However,
once the jet is substantially bent from its original direction by the
putative SNR wind, the pulsar's motion with respect to the SNR matter can bend it
further, so that the jet can become directed backwards (towards southeast).
A hint of such a bend is indeed seen in Figure 1. Moreover, the relativistic
particles blown off the jet by the SNR wind are being picked up by the head wind
(in the pulsar's reference frame) and dragged in the direction opposite
to the pulsar's proper motion, feeding the diffuse nebula. Such a picture
is supported by the spectral slope of the diffuse emission, $\Gamma \approx 1.5$,
which is softer than the emission of the brighter outer jet ($\Gamma\approx
1.3$), but harder than the emission of the PWN shell ($\Gamma\approx 1.65$).
Thus, we can conclude that the outermost observable part of outer jet is
likely bent by the combined action of the SNR wind and the pulsar's proper
motion, and the outer jet finally bends backward and get diffused southwest
of the pulsar.
\subsection{Loop-like structures and blobs: Instabilities in pinched flow?}
So far we have discussed only the bending of the jet's ``end''. Explaining
more extreme bends in the brighter part of the outer jet
(e.g., the hook-like or loop-like structures in panels 1, 2, 4, and 13)
by the wind action is more problematic. Although we see that these bends
are convex in the direction of the SNR wind suggested above,
the relatively small size (and strong curvature) of these structures would
require a strong nonuniformity and very high velocities of the wind. Even a stronger
argument is the observed variability of the bent structures,
associated with almost relativistic velocities. Therefore, we have to invoke
another mechanism to explain these features, not related to external winds.
A natural explanation is the kink instability of a magnetically confined,
pinched flow. Estimates of the instability growth times and wavelengths
depend on the model of such a flow, particularly the distribution of
currents and topology of the magnetic field. The flow model is also
tightly connected with the interpretation
of the whole PWN, particularly the inner jets and arcs.
Analysis of these models
is beyond the scope of the present paper. Therefore, we only briefly
mention that the (outer) jet can be modeled as a plasma beam carrying
a charge current of $\sim 10^{32}$ $e$ s$^{-1}$ ($\sim \dot{N}_{\rm jet}$
--- see \S3.2),
self-confined by a predominantly toroidal magnetic field (Z pinch).
A similar model for pulsar jets was suggested by Benford (1984),
while collimation and confinment of AGN jets by magnetic field was
reviewed
by Begelman, Blandford \& Rees (1984)
and other authors.
Growth times of MHD instabilities in such a flow are proportional to the
Alfven crossing time, $\tau_A \sim r_{\rm jet}/v_A$ (e.g., Begelman 1998).
Using the expression for the Alfven velocity in the ultrarelativistic
plasma, $v_A=cB [4\pi(w_{\rm rel} +p_{\rm rel}) +B^2]^{-1/2}$
(Akhiezer et al.\ 1975), we obtain $v_A=(3/5)^{1/2} c\simeq 0.77 c$
at equipartition between the magnetic and particle energy densities,
which gives $\tau_A\sim 10$ days.
This time is comparable to the
time scales of strong bendings ($\la 30$ days), which can be associated
with the kink instabilities, and time scales of blob brightening (about
a week), which can be associated with the sausage (neck) instabilities.
\section{Summary and Conclusions}
The multiple {\sl Chandra} observations of the Vela PWN have allowed us
to investigate the dynamical outer jet and discover a dimmer outer
counter-jet outside the bright PWN.
Our main results can be summarized as follows.
\medskip

1. The outer jet extends up to about 0.4--0.5 light years from the pulsar
in the sky plane
along the direction of the pulsar's proper motion. Its shape and
brightness are variable on time scales of days to weeks. The brightness is
nonuniform along the jet, with brighter blobs moving away from the pulsar
with apparent subrelativistic speeds. The variations observed suggest
typical flow velocities of 0.3--0.7 of the speed of light.

2. The outer jet is, on average, a factor of 7 brighter than the outer counter-jet.
If the outer jet and outer counter-jet are intrinsically similar but
streaming in opposite directions, then the difference in brightness means that
the outer jet is approaching at an angle of $30^\circ$--$70^\circ$
to the line-of sight
while the outer counter-jet is receding. Such an orientation apparently
contradicts to the previously suggested models of the inner jets and the bright
arcs.

3. The synchrotron interpretation of the  hard ($\Gamma \approx 1.3$)
power-law spectrum of the outer
jet requires highly relativistic electrons or positrons, with energies of up
to $\ga 200$ TeV, and a typical
magnetic field of about 100 $\mu$G. The outer jet's
spectrum is perhaps slightly softer than those of the inner jet and counter-jet,
but it  does not change appreciably along the outer jet. If the outer jet
is a mildly relativistic outflow of an ultrarelativistic electron/positron
plasma, the energy injection rate is $\sim 10^{34}\, {\rm erg\, s}^{-1}
\sim 10^{-3} \dot{E} \sim 10^3 L_{\rm x,jet}$.

4. Outside the bright PWN,
there is an asymmetric, dim outer diffuse nebula that is substantially brighter
southwest of the jet/counter-jet line. Its spectrum is softer than that of the
outer jet, but it is harder than the spectrum
of the brighter PWN shell.
It is possible that the X-ray emitting particles in the
dim nebula are supplied through the outer jet, whose
 end part turns southwest
with respect
to its average (northwest) direction in the sky plane.
Such a turn can be caused by a northeast SNR wind with a speed of a few times
10 km s$^{-1}$, which also helps feed the dim nebula.

5. The width of the outer jet, $\sim 3\times 10^{16}$ cm, remains approximately
the same along the jet in different observations, including those which
show strong bends. This suggests an efficient confinement mechanism, perhaps
associated with magnetic fields generated by electric currents in the pinched
jet. The current required, $\sim 10^{32}\, e\, {\rm s}^{-1}\sim 10^{12}\, {\rm amp}$,
is an order of magnitude lower than the Goldreich-Julian current in the pulsar
magnetosphere. The bright blobs and strong bends could be caused by the sausage
and kink instabilities, respectively, in such a pinched jet.

\medskip\noindent
The excellent resolution of the {\sl Chandra} telescope
and high sensitivity of its detectors allowed us to obtain the spectacular
pictures of the Vela PWN, including the faint outer jet, and, in
particular, to prove the highly anisotropic and dynamical
nature of the pulsar outflows.
However, as has been often the case with new high-quality data, our results
raise new questions and put under doubt the previous simplistic interpretations
of PWNe in general and the Vela PWN in particular. The most unclear issue
is the interpretation of the complicated morphology of the Vela PWN,
particularly the relationship of the outer jet and outer counter-jet
with their inner counterparts, the true orientation of the jets, and
the actual topology of its arcs. We do not even know the actual three-dimentional
orientations of the inner and outer jets and counter-jets, nor do we understand
the cause of their different brightness.
We can only guess about the nature of the bright blobs moving along the outer
jet, the mechanism(s) of the jet confinement, and the origin of the jet bendings.
At least some of these issues can be clarified by a series
of deeper {\sl Chandra} observations of the Vela PWN, taken with intervals of
a few days, the now-established time scale of the PWN variations. In particular,
such observations would allow one to search for the spectral changes along
the outer jet and check if the spectra of the blobs are different from those
of the rest of the jet. They could also help find blobs in the very dim
outer counter-jet and measure their velocities, which would provide a clue
to the geometry of the system.

The interpretation of the Vela PWN would be much easier if it
were detected at other wavelengths, outside the X-ray range. So far, contrary
to the much better studied Crab PWN, the Vela PWN has not been detected in
the optical, mainly because of numerous relatively bright field stars and
SNR filaments in the pulsar vicinity (Mignani et al.\ 2003).
To get rid of their light and detect
the PWN or put a stringent upper limit on its brightness, the field must
be observed in polarized light. It would also be very important to obtain
a deep radio image
the field around the Vela pulsar with arcsecond resolution (cf.\ Dodson
et al.\ 2003b).  Both the optical
and radio observations would provide estimates on the lower frequencies
of the synchrotron spectra (hence, lower energies of relativistic electrons),
the critical parameters for evaluating the magnetic fields and the energetics
of the observed PWN elements. Measuring polarizations of the optical and radio
emission would be crucial to establish the directions of the magnetic field
in the PWN elements and understand their nature.

\acknowledgements
We thank Leisa Townsley and George Chartas for the advice on ACIS data reduction
and Bing Zhang for useful discussions of pulsar physics.
Support for this work was provided by the NASA
through Chandra Awards GO1-2071X and GO2-3091X,
issued by the Chandra X-Ray Observatory Center,
which is operated by the Smithsonian Astrophysical Observatory
for and on behalf of NASA under contract NAS8-39073.
This work was also partially supported by NASA grant NAG5-10865.
\newpage



\begin{thebibliography}{}
\bibitem[1]{1}
Aschenbach, B., \& Brinkmann, W. 1975, A\&A, 41, 147
\bibitem[2]{2}
Akhiezer, A.\ I., Akhiezer, I.\ A., Polovin, R.\ V., Sitenko, A.\ G.,
\& Stepanov, K.\ N. 1975, Plasma Electrodynamics, (Oxford: Pergamon), p.112
\bibitem[3]{3}
Begelman, M.\ C. 1998, ApJ, 493, 291
\bibitem[4]{4}
Begelman, M.\ C., Blandford, R.\ D., Rees, M.\ J. 1984, Rev.\ Mod.\ Phys., 56, 255
\bibitem[5]{5}
Benford, G. 1984, ApJ, 282, 154
\bibitem[6]{6}
Caraveo, P.\ A., De Luca, A., Mignani, R.\ P., \& Bignami, G.\ F. 2001,
ApJ, 561, 930
\bibitem[7]{7}
Dodson, R., Legge, D., Reynolds, J.\ E., \& McCulloch, P.\ M. 2003a,
ApJ, submitted (astro-ph/0302374)
\bibitem[8]{8}
Dodson, R., Lewis, D., McConnel, D., \& Deshpande, A.\ A. 2003b,
MNRAS, submitted (astro-ph/0302373)
\bibitem[9]{9}
Gaensler, B.\ M., Arons, J., Caspi, V.\ M., Pivovaroff, M.\ J., Kawai, N.,
\& Tamura, K. 2002, ApJ, 569, 878
\bibitem[10]{10}
Gallant, Y.\ A., \& Arons, J. 1994, ApJ, 435, 230
\bibitem[11]{11}
Ginzburg, V.\ L., \& Syrovatskii, S.\ I. 1965, ARA\&A, 3, 297 [GS65]
\bibitem[12]{12}
Harding, A.\ K., Muslimov, A.\ G., \& Zhang, B. 2002, ApJ, 576, 366
\bibitem[13]{13}
Harnden, F.\ R., Grant, P.\ D., Seward, F.\ D., \& Kahn, S.\ M. 1985,
ApJ, 299, 828
\bibitem[14]{14}
Helfand, D.\ J., Gotthelf, E.\ V., \& Halpern, J.\ P. 2001, ApJ, 556, 380
\bibitem[15]{15}
Hester, J.\ J., Mori, K., Burrows, D., et al. 2002, ApJ, 577, L49
\bibitem[16]{16}
Hibschman, J., \& Arons, J. 2001, ApJ, 560, 871
\bibitem[17]{17}
Kargaltsev, O., Pavlov, G.\ G., Sanwal, D., \& Garmire, G.\ P. 2002,
in Neutron Stars in Supernova Remnants, eds.\ P.\. O.\ Slane \&
B.\ M.\ Gaensler, ASP Conf.\ Ser., v.271 (San Francisco: ASP), 181
\bibitem[18]{18}
Leahy, J.\ P. 1991, in Beams and Jets in Astrophysics, ed.\ P.\ A.\ Hughes
(Cambridge: Cambridge Univ.\ Press), 147
\bibitem[19]{19}
Lewis, D., Dodson, R., McConnel, D., \& Deshpande, A. 2002,
in Neutron Stars in Supernova Remnants, eds.\ P.\. O.\ Slane \&
B.\ M.\ Gaensler, ASP Conf.\ Ser., v.271 (San Francisco: ASP), 191
\bibitem[20]{20}
Markwardt, C.\ B., \& \"Ogelman, H. 1998, Mem.\ Soc.\ Astron.\ Italiana,
69, 927
\bibitem[21]{21}
Mignani, R.\ P., De Luca, A., Kargaltsev, O., Pavlov, G.\ G., Zaggia, S.,
Caraveo, P.\ A., \& Becker, W.
2003, ApJ, submitted
\bibitem[22]{22}
Mori, K., Hester, J.\ J., Burrows, D.\ N., Pavlov, G.\ G., \& Tsunemi, H. 2002,
in Neutron Stars in Supernova Remnants, ASP Conf.\ Ser., v.\ 271, eds.\
                P.\ O.\ Slane \& B.\ M.\ Gaensler (San Francisco: ASP), 157
\bibitem[23]{23}
Pacholczyk, A.\ G. 1970, Radio Astrophysics (San Francisco: Freeman)
\bibitem[24]{24}
Pavlov, G.\ G. 2000, AAS HEAD Meeting 32, \#07.05
\bibitem[25]{25}
Pavlov, G.\ G., Sanwal, D., Garmire, G.\ P., Zavlin, V.\ E., Burwitz, V.,
\& Dodson, R.\ G. 2000, AAS Meeting 196, \#37.04
\bibitem[26]{26}
Pavlov, G.\ G., Kargaltsev, O.\ Y., Sanwal, D., \& Garmire, G.\ P. 2001a,
ApJ, 554, L189
\bibitem[27]{27}
Pavlov, G.\ G., Zavlin, V.\ E., Sanwal, D., Burwitz, V,. \& Garmire, G.\ P.
2001b, ApJ, 552, L129
\bibitem[28]{28}
Plucinsky, P.\ P., et al. 2002, in Astronomical Telescopes and Instrumentation 2002,
eds.\ J.\ E.\ Tr\"emper \& H.\ D.\ Tananbaum, SPIE Conf. Proc., in press
(astro-ph/0209161)
\bibitem[29]{29}
Radhakrishnan, V., \& Deshpande, A. 2000, A\&A, 379, 551
\bibitem[30]{30}
Townsley, L.\ K., Broos, P.\ S., Garmire, G.\ P., \& Nousek, J.\ A. 2000, ApJ, 534, L139
\bibitem[31]{31}
Zhang, L., \& Cheng, K.\ S. 1997, ApJ, 487, 370
\end{thebibliography}
\end{document}